\documentclass[twocolumn,prd,superscriptaddress,showpacs,floatfix,%
preprintnumbers,nofootinbib]{revtex4}

\usepackage{bm}

\begin{document}

\title{Time Structure of Ultra-High Energy Cosmic Ray Sources and Consequences
for Multi-messenger Signatures}
\author{G{\"u}nter Sigl}

\affiliation{II. Institut f\"ur theoretische Physik, Universit\"at Hamburg,
Luruper Chaussee 149, D-22761 Hamburg, Germany}
\affiliation{APC~\footnote{UMR 7164 (CNRS, Universit\'e Paris 7,
CEA, Observatoire de Paris)} (AstroParticules et Cosmologie),
10, rue Alice Domon et L\'eonie Duquet, 75205 Paris Cedex 13, France}

\begin{abstract}
The latest results on the sky distribution of ultra-high energy cosmic ray
sources have consequences for their nature and time structure. If the
sources accelerate predominantly nuclei of atomic number $A$ and charge $Z$
and emit continuously,
their luminosity in cosmic rays above $\simeq6\times10^{19}\,$eV can be no
more than a fraction of $\simeq5\times10^{-4}\,Z^{-2}$ of their total power output.
Such sources could produce a diffuse neutrino flux that gives rise to several
events per year in neutrino telescopes of km$^3$ size. Continuously emitting
sources should be easily visible in photons below $\sim100\,$GeV, but not in TeV
$\gamma-$rays which are
likely absorbed within the source. For episodic sources that are beamed
by a Lorentz factor $\Gamma$, the bursts or flares have to last at least
$\simeq0.1\,\Gamma^{-4}\,A^{-4}\,$yr. A considerable fraction of the flare
luminosity could go into highest energy cosmic rays, in which case
the rate of flares per source has to be less than
$\simeq5\times10^{-3}\,\Gamma^4\,A^4\,Z^2\,{\rm yr}^{-1}$. Episodic sources
should have detectable variability both at GLAST and TeV energies, but
neutrino fluxes may be hard to detect.
\end{abstract}


\maketitle

\section{Introduction}
The sources of ultra-high energy cosmic rays (UHECR) above
the ``GZK threshold''~\cite{gzk} around $\simeq6\times10^{19}\,$eV
are still unknown. Recently, an important step forward has been
made by the Pierre Auger Observatory which has revealed a correlation of the
arrival directions of UHECR above $\simeq6\times10^{19}\,$eV with
the nearby cosmological large scale structure as mapped out by the
distribution of active galactic nuclei (AGNs)~\cite{Cronin:2007zz}.
At least if UHECR deflection in large scale cosmic magnetic fields is moderate,
this requires a certain minimal density of sources within the "GZK horizon"
of about 75 Mpc. At the same time the observed spectrum normalizes the
required injection power per volume. Together, these two numbers imply an upper
limit on the time averaged UHECR injection power per source. Comparing
this with the minimal power that needs to be dissipated in order to
produce UHECR up to $\sim10^{20}\,$eV, this allows to constrain the
time structure of the UHECR emission, in particular continuously emitting
versus episodic sources. This can also have some implications on how the dissipated
total power may be distributed between the cosmic ray, photon, and neutrino
channels. The latter can be important for a multimessenger
study of UHECR sources. In the present paper we attempt to work out
these constraints in a largely model-independent way, including their
dependence on the type of nuclei that are predominantly accelerated.

The remainder of this paper is structured as follows. In Sect.~II we
develop general requirements on the individual sources.
In Sect.~III and~IV we consider continuously emitting and episodic sources,
respectively. In Sect.~V we discuss Centaurus A as a potential UHECR source
and we conclude in Sect.~VI. We will use the units in which $c=1$ throughout.

\section{Requirements on Individual Sources}
Accelerating particles of charge $eZ$ to an energy $E_{\rm max}$ requires
an induction ${\cal E} \gtrsim E_{\rm max}/(eZ)$. With $Z_0\simeq100\,\Omega$
the vacuum impedance, this requires dissipation of a minimal power
of~\cite{Lovelace,Blandford:1999hi}
\begin{equation}\label{eq:Lmin}
  L_{\rm min}\simeq\frac{{\cal E}^2}{Z_0}\simeq 10^{45}\,Z^{-2}\,
  \left(\frac{E_{\rm max}}{10^{20}\,{\rm eV}}\right)^2\,
  {\rm erg}\,{\rm s}^{-1}\,.
\end{equation}
We stress that this minimal power can be less by factors of order unity
in specific geometrical circumstances, such as magnetic fields connecting
accretion disks with jets in AGNs~\cite{biermann}. However, given other,
larger uncertainties such as the chemical composition of UHECRs, we can
ignore such details in the present work.

The "Poynting" luminosity Eq.~(\ref{eq:Lmin}) can also be obtained from the expression
$L_{\rm min}\sim\Gamma^2(BR)^2$ where $\Gamma$ is the beaming factor
of the accelerating region and the product of the size $R$ and magnetic
field strength $B$ of the acceleration region is given by the
"Hillas criterium"~\cite{hillas-araa} which states that the Larmor radius
$r_{\rm L}=E_{\rm max}/(\Gamma eZB)$ should be smaller than $R$,
\begin{equation}\label{eq:Hillas}
  \left(\frac{B}{{\rm G}}\right)\left(\frac{R}{{\rm cm}}\right)
  \gtrsim 3\times10^{17}\,\Gamma^{-1}\,
  \left(\frac{E_{\rm max}}{Z10^{20}\,{\rm eV}}\right)\,.
\end{equation}
In the following we denote cosmic rays above $6\times10^{19}\,$eV as
ultra-high energy cosmic rays (UHECR) and we take $E_{\rm max}\simeq10^{20}\,$eV
as the benchmark for their typical production energy within the sources.
Any source producing UHECR up to energy $E_{\rm max}$ at a given time
has to have a total power output of at least the Poynting luminosity
Eq.~(\ref{eq:Lmin}). Note that this is comparable to the Eddington luminosity
$L_{\rm Edd}(M)=1.3\times10^{38}(M/M_\odot)\,{\rm erg}\,{\rm s}^{-1}$
of a massive black hole of mass $M$ in the centers of active galaxies.
A considerable part $L_\gamma$ of that
power is presumably electromagnetic and thus emitted in photons. We now
assume that electromagnetic power is produced in the same area of size
$R$ in which UHECR are accelerated. Denoting the characteristic photon energy by
$\varepsilon$, the optical depth for pion production on such photons
by accelerated protons with an energy above the photo-pion threshold,
$E\gtrsim6.8\times10^{16}(\varepsilon/{\rm eV})^{-1}\,$eV, is given by
\begin{eqnarray}\label{eq:tau_pgamma}
  \tau_{p\gamma}&\simeq&\sigma_{p\gamma}n_\gamma R\simeq
  \frac{\sigma_{p\gamma}L_\gamma}{4\pi R\varepsilon}\\
  &\simeq&
  0.15\left(\frac{L_\gamma}{10^{45}\,{\rm erg}\,{\rm s}^{-1}}\right)\,
  \left(\frac{R}{{\rm pc}}\right)^{-1}\,
  \left(\frac{\varepsilon}{{\rm eV}}\right)^{-1}\,,\nonumber
\end{eqnarray}
where we have used $\sigma_{p\gamma}\simeq300\mu$barn around the threshold
for pion production.
Note that $R\sim1\,$pc is the typical size of an accretion disk around a
supermassive black holes at the centers of AGNs which
is determined by the "sphere of influence"
$\sim2G_{\rm N}M/v_s^2\sim2\,(M/10^7\,M_\odot)(v_s/200\,{\rm km}\,{\rm s}^{-1})^{-2}\,$pc where
$G_{\rm N}$ is Newton's constant and $v_s$ is the velocity dispersion of the
stars in the host galaxy~\cite{agn-disks}.
The optical depth for photo-disintegration of primary nuclei is comparable to
Eq.~(\ref{eq:tau_pgamma}).
If it is significantly larger than unity, most nuclei would be disintegrated
before leaving the source and the maximal energy would have to be
$E_{\rm max}\gtrsim A\,10^{20}\,$eV in order for UHECR to arrive at
Earth with energies up to $10^{20}\,$eV.

On the other hand, the optical depth for hadronic interactions of accelerated
protons and nuclei with the surrounding bulk matter of hadronic mass $M_{\rm bulk}$
extending over a characteristic scale $R_{\rm bulk}\gtrsim R$ can be written as
\begin{eqnarray}\label{eq:tau_pp}
  \tau_{pp}&\simeq&\sigma_{pp}n_p R_{\rm bulk}\simeq
  \frac{\sigma_{pp}M_{\rm bulk}}{R_{\rm bulk}^2 m_N}\\
  &\simeq&100\left(\frac{M_{\rm bulk}}{10^7\,M_\odot}\right)\,
  \left(\frac{R_{\rm bulk}}{{\rm pc}}\right)^{-2}\,\,,\nonumber
\end{eqnarray}
where we have estimated the nucleon density by
$n_p\sim M_{\rm bulk}/R_{\rm bulk}^3$. Note that
the mass of AGN accretion disks is roughly comparable to the mass of the
central supermassive black hole~\cite{agn-disks} whose typical mass
is $10^{7-8}\,M_\odot$. Eq.~(\ref{eq:tau_pp}) is only a rough
estimate because the details will depend on the geometry, for example spherical versus
disk-like accretion. Since the bolometric luminosities of most AGNs
are $\ll10^{47}\,{\rm erg}\,{\rm s}^{-1}$, a comparison of Eqs.~(\ref{eq:tau_pgamma})
and~(\ref{eq:tau_pp}) suggests that hadronic interactions dominate over photo-hadronic
interactions in the cores of AGNs.

Pionic and photo-hadronic processes will produce secondary $\gamma-$rays
and neutrinos. The optical depth for photons of energy above the pair
production threshold,
$E\gtrsim m_e^2/\varepsilon\simeq0.26(\varepsilon/{\rm eV})^{-1}\,$TeV,
can be estimated as
\begin{eqnarray}\label{eq:tau_gg}
  \tau_{\gamma\gamma}&\simeq&\sigma_{\rm T}n_\gamma R\simeq
  \frac{\sigma_{\rm T}L_\gamma}{4\pi R\varepsilon}\\
  &\simeq&
  300\left(\frac{L_\gamma}{10^{45}\,{\rm erg}\,{\rm s}^{-1}}\right)\,
  \left(\frac{R}{{\rm pc}}\right)^{-1}\,
  \left(\frac{\varepsilon}{{\rm eV}}\right)^{-1}\,,\nonumber
\end{eqnarray}
where $\sigma_{\rm T}\simeq0.6\,$barn is the Thomson cross section.

We now deduce some requirements on the size $R$ of the accelerating region.
We will also take into account a possible beaming factor $\Gamma$ such
that $B$ and $R$ and other length scales are measured in the comoving frame,
whereas luminosities and the energy $E_{\rm max}$ refer to the observer frame.
The synchrotron loss length for a nucleus of atomic number $A$ and charge $Z$
in a magnetic field of strength $B$ is
\begin{equation}\label{eq:l_synch}
  l_{\rm synch} \simeq0.43\,\Gamma\,A^4\,Z^{-2}\,
  \left(\frac{E_{\rm max}}{10^{20}\ {\rm eV}}\right)^{-1}
  \left(\frac{B}{{\rm G}}\right)^{-2}\,{\rm pc}\,,
\end{equation}
and the Larmor radius can be written as
\begin{equation}\label{eq:r_Larmor}
  r_{\rm L} \simeq 0.1\,\Gamma^{-1}\,Z^{-1}\,
  \left(\frac{E_{\rm max}}{10^{20}\ {\rm eV}}\right)
  \left(\frac{B}{{\rm G}}\right)^{-1}\,{\rm pc}\,.
\end{equation}
Since any energy loss time must be longer than the acceleration time
which itself is larger than the Larmor radius, one has the condition
$l_{\rm synch}\gtrsim r_{\rm L}$ which gives an upper limit on the
magnetic field strength,
\begin{equation}\label{eq:B_cond}
  B \lesssim 4.3\,\Gamma^2\,A^4\,Z^{-1}\,
  \left(\frac{E_{\rm max}}{10^{20}\ {\rm eV}}\right)^{-2}
  \,{\rm G}\,.
\end{equation}
Together with Eq.~(\ref{eq:r_Larmor}) this results in a lower limit on
the Larmor radius,
\begin{equation}\label{eq:r_L_cond}
  r_{\rm L} \gtrsim 2.3\times10^{-2}\,\Gamma^{-3}\,A^{-4}\,
  \left(\frac{E_{\rm max}}{10^{20}\ {\rm eV}}\right)^3
  \,{\rm pc}\,.
\end{equation}
In case of AGN sources, for example, this is certainly consistent
with a size $R\sim1\,$pc for the typical size of an accretion disk.
Acceleration could thus occur in small parts of such accretion disks.

\section{Continuously emitting Sources}
Assuming at most moderate
deflection in intergalactic space, the number of arrival directions
observed by the Pierre Auger Observatory and other experiments
implies a lower limit on the source
density~\cite{Kachelriess:2004pc,Farrar:2008ex},
\begin{equation}\label{eq:ns}
  n_s\gtrsim3\times10^{-5}\,{\rm Mpc}^{-3}\,.
\end{equation}
The UHECR flux observed by the Pierre Auger observatory is~\cite{Roth:2007in}
\begin{equation}\label{eq:diff_flux}
  \frac{dN_{\rm CR}}{dE}(E\simeq6\times10^{19}\,{\rm eV})\simeq
  6\times10^{-40}\,{\rm cm}^{-2}\,{\rm sr}^{-1}\,{\rm s}^{-1}\,{\rm eV}^{-1}\,,
\end{equation}
which corresponds to a power per volume of~\cite{Farrar:2008ex}
\begin{equation}\label{eq:Quhecr}
  Q_{\rm UHE}\sim1.3\times10^{37}\,{\rm erg}\,{\rm Mpc}^{-3}\,
  {\rm s}^{-1}\,.
\end{equation}
Comparing Eqs.~(\ref{eq:ns}) and~(\ref{eq:Quhecr}) implies for
the time-averaged UHECR luminosity per source
\begin{equation}\label{eq:Luhecr}
  \overline{L_{\rm UHE}}\lesssim4\times10^{41}\,{\rm erg}\,{\rm s}^{-1}\,.
\end{equation}
This is much smaller than the instantaneous total luminosity required by
Eq.~(\ref{eq:Lmin}).

If UHECR sources emit continuously, Eqs.~(\ref{eq:Lmin}) and~(\ref{eq:Luhecr})
imply that these sources must emit at least $\simeq2000\,Z^{-2}$ times more
energy in channels
other than UHECR. This is consistent with the fact that at redshift zero
an average AGN in an active state has a bolometric luminosity of
$\simeq5\times10^{44}\,{\rm erg}\,{\rm s}^{-1}$, comparable to Eq.~(\ref{eq:Lmin}),
and the volume emissivity
is $\simeq3\times10^{40}\,{\rm erg}\,{\rm Mpc}^{-3}\,{\rm s}^{-1}$, a factor
of a few thousand larger than Eq.~(\ref{eq:Quhecr})~\cite{Marconi:2003tg,Ueda:2003yx}.
The average AGN bolometric luminosity and volume emissivity corresponds to a
density of "typical" AGNs of $\simeq6\times10^{-5}\,{\rm Mpc}^{-3}$,
consistent with Eq.~(\ref{eq:ns}).

As a result, if sources emit continuously and the total power is distributed roughly
equally between hadronic cosmic rays and electromagnetic power, the cosmic ray injection
spectrum could extend down to $\lesssim10^{17}\,$eV with a rather steep spectrum
$\propto E^{-\alpha}$, $\alpha\lesssim2.7$. Using Eq.~(\ref{eq:Quhecr}), this can
be written as
\begin{equation}\label{eq:j_inj}
  \frac{d\dot{n}_{\rm CR}}{dE}(E)\simeq(\alpha-2)\frac{Q_{\rm UHE}}{(10^{20}\,{\rm eV})^2}
  \left(\frac{E}{10^{20}\,{\rm eV}}\right)^{-\alpha}\,.
\end{equation}
Furthermore, Eq.~(\ref{eq:tau_pp}) suggests
that the optical depth for hadronic interactions can be of order unity and
thus a considerable part of that cosmic ray flux could be transformed to neutrinos
with energies $\sim10^{17}\,$eV. Following Ref.~\cite{Halzen:2008vz}, for proton
primaries, $Z=1$, we can write for the production rate per volume of neutrinos
\begin{equation}\label{eq:j_inj_nu}
  \frac{d\dot{n}_\nu}{dE}(E)\simeq\frac{2f}{3x_\nu}
  \frac{d\dot{n}_{\rm CR}}{dE}(E/x_\nu)\,,
\end{equation}
where $x_\nu\simeq0.05$ is the average neutrino energy in units of
the parent cosmic ray energy and $f=e^\tau-1$ is the ratio of number
of cosmic rays interacting within the source to cosmic rays leaving the source. If
the cosmic ray injection spectrum $\propto E^{-\alpha}$
extends down to $E_{\rm min}$ without break, $f$ is limited by
\begin{equation}\label{eq:f}
  f\left(\frac{E_{\rm min}}{10^{20}\,{\rm eV}}\right)^{2-\alpha}\lesssim
  2\times10^3\,Z^{-2}\frac{L_{\rm tot}}{L_{\rm min}}\,,
\end{equation}
where $L_{\rm tot}$ is the total luminosity and $L_{\rm min}$ is given
by Eq.~(\ref{eq:Lmin}). This condition just results from comparing the
total emissivity with the output in UHECR and neutrinos and would be saturated if the
total output would be dominated by neutrinos in case of "hidden sources".
Since neutrinos do not interact during propagation and ignoring
redshift evolution, we can estimate the all-flavor diffuse neutrino flux as
\begin{equation}\label{eq:j_diff_nu}
  j_\nu^{\rm diff}(E)\simeq\frac{1}{4\pi H_0}\frac{d\dot{n}_\nu}{dE}(E)\,,
\end{equation}
where $H_0=100\,h{\rm km}\,{\rm s}^{-1}\,{\rm Mpc}^{-1}$ is the Hubble constant
with $h\simeq0.72$. Putting together Eqs.~(\ref{eq:j_inj}), (\ref{eq:j_inj_nu})
and~(\ref{eq:j_diff_nu}), we obtain
\begin{eqnarray}\label{eq:j_diff_nu2}
  E^2j_\nu^{\rm diff}(E)&\simeq&190\,x_\nu^{\alpha-1}(\alpha-2)f\\
  &&\times\left(\frac{E}{10^{20}\,{\rm eV}}\right)^{2-\alpha}\,
  {\rm eV}\,{\rm cm}^{-2}\,{\rm s}^{-1}\,{\rm sr}^{-1}\,.\nonumber
\end{eqnarray}

If the ankle marks the transition from galactic to extragalactic cosmic rays,
the injection spectral index of the latter has to be $\alpha\simeq2.2$, especially
if heavier nuclei are accelerated~\cite{Allard:2005cx}. The secondary neutrino
spectrum would then extend down to at least $\simeq10^{17}\,$eV and
Eq.~(\ref{eq:j_diff_nu2}) implies
\begin{equation}\label{eq:j_diff_nu3}
  E^2j_\nu^{\rm diff}(E)\simeq4.2\,f
  \left(\frac{E}{10^{17}\,{\rm eV}}\right)^{-0.2}\,
  {\rm eV}\,{\rm cm}^{-2}\,{\rm s}^{-1}\,{\rm sr}^{-1}\,.
\end{equation}
In contrast, if the ankle is due to pair production of extragalactic protons,
then one needs $\alpha\simeq2.6$~\cite{Berezinsky:2005cq}. The secondary neutrino
spectrum would then extend down to at least $\simeq10^{16}\,$eV and
Eq.~(\ref{eq:j_diff_nu2}) implies
\begin{equation}\label{eq:j_diff_nu4}
  E^2j_\nu^{\rm diff}(E)\simeq240\,f
  \left(\frac{E}{10^{16}\,{\rm eV}}\right)^{-0.6}\,
  {\rm eV}\,{\rm cm}^{-2}\,{\rm s}^{-1}\,{\rm sr}^{-1}\,.
\end{equation}

We now compute the neutrino event rates in kilometer scale neutrino
observatories for these two scenarios. Using the neutrino-nucleon
cross section $\sigma_{\nu N}\simeq1.9\times10^{-33}(E/10^{16}\,{\rm eV})^{0.363}\,{\rm cm}^2$
for $10^{16}\,{\rm eV}\lesssim E\lesssim10^{21}\,$eV~\cite{Gandhi:1998ri},
we obtain the rate
\begin{eqnarray}\label{eq:nurate}
  R_\nu&\sim&
  \sigma_{\nu N}(E)2\pi E j_\nu^{\rm diff}(E)n_N V_{\rm eff}\\
  &\sim&2.3\left(\frac{E}{10^{16}\,{\rm eV}}\right)^{-0.637}
  \left(\frac{V_{\rm eff}}{{\rm km}^3}\right)
  \nonumber\\
  &&\times\left(\frac{E^2 j_\nu^{\rm diff}(E)}
  {100\,{\rm eV}{\rm cm}^{-2}{\rm sr}^{-1}{\rm s}^{-1}}\right)
  \,{\rm yr}^{-1}\,,
  \nonumber
\end{eqnarray}
where $n_N\simeq6\times10^{23}\,{\rm cm}^{-3}$ is the nucleon
density in water/ice and $V_{\rm eff}$ the effective detection volume.
The scenario $\alpha=2.2$ with neutrino flux down to $10^{17}\,$eV would
give $\simeq2\times10^{-2}\,f\,{\rm yr}^{-1}\,{\rm km}^{-3}
\lesssim20\,Z^{-2}(L_{\rm tot}/L_{\rm min})\,{\rm yr}^{-1}\,{\rm km}^{-3}$,
where we have used Eq.~(\ref{eq:f}) for $f$. The scenario
$\alpha=2.6$ with neutrino flux down to $10^{16}\,$eV would
give $\simeq5.5\,f\,{\rm yr}^{-1}\,{\rm km}^{-3}
\lesssim180\,Z^{-2}(L_{\rm tot}/L_{\rm min})\,{\rm yr}^{-1}\,{\rm km}^{-3}$.

TeV $\gamma-$rays would mostly be absorbed within the source due to Eq.~(\ref{eq:tau_gg}).
As opposed to Ref.~\cite{Halzen:2008vz}, we therefore do not get a constraint
from the non-observation of AGNs at TeV energies in this scenario.
In contrast, X-rays and GeV $\gamma-$rays could leave the source. Individual
sources should be visible by X-ray telescopes and by GLAST. In fact,
EGRET has seen a diffuse flux~\cite{egret}
which constrains the neutrino flux because a comparable amount of energy goes
into photons and neutrinos in primary cosmic ray interactions,
\begin{equation}\label{eq:EGRET}
  E^2j_\nu^{\rm diff}(E)\lesssim10^{3}\,
  {\rm eV}\,{\rm cm}^{-2}\,{\rm s}^{-1}\,{\rm sr}^{-1}\,.
\end{equation}
Eqs.~(\ref{eq:j_diff_nu3}) and~(\ref{eq:j_diff_nu4}) satisfy this
limit except for sources deeply in the hidden regime, $f\gg1$.

\section{Flaring Sources}
If the sources flare on a typical time scale $\delta t$ in the observer
frame, the corresponding life
time of the burst in the comoving frame, $\Gamma\delta t$ must be
larger than the comoving acceleration time scale which itself is larger
than the Larmor radius, thus with Eq.~(\ref{eq:r_L_cond}) we have
\begin{equation}\label{eq:delta_t}
  \delta t \gtrsim 0.1\,\Gamma^{-4}\,A^{-4}\,
  \left(\frac{E_{\rm max}}{10^{20}\ {\rm eV}}\right)^3
  \,{\rm yr}\,.
\end{equation}
This is consistent with the variabilities observed for AGNs which are
observed at time scales down to $\sim60\,$s~\cite{Albert:2007zd}, provided
that $\Gamma\gtrsim20$ and/or predominantly heavier nuclei are accelerated.
The time scale is also consistent with $\gamma-$ray bursts which can easily have
Lorentz factors $\Gamma\gtrsim20$~\cite{Waxman:1995vg}.

If we denote the rate and UHECR luminosity of typical flares by $R_{\rm f}$ and
$L_{\rm UHE}$, respectively, we can write for the time averaged UHECR power,
Eq.~(\ref{eq:Luhecr}),
\begin{equation}\label{eq:flare}
  R_{\rm f}\delta t L_{\rm UHE}\sim\overline{L_{\rm UHE}}
  \lesssim4\times10^{41}\,{\rm erg}\,{\rm s}^{-1}\,.
\end{equation}
This means that the fraction of time a typical intermittent source emits
the typical instantaneous UHECR luminosity $L_{\rm UHE}$, also called
{\it duty factor}, is given by
\begin{equation}\label{eq:duty}
  {\cal D}\equiv R_{\rm f}\delta t\lesssim4\times10^{-4}\,Z^2\,
  \left(\frac{L_{\rm min}}{L_{\rm UHE}}\right)\,
  \left(\frac{E_{\rm max}}{10^{20}\ {\rm eV}}\right)^{-2}\,.
\end{equation}
Eqs.~(\ref{eq:flare}) and~(\ref{eq:duty})
imply that a considerable fraction of the minimal flaring luminosity
Eq.~(\ref{eq:Lmin}) can go into UHECR, $L_{\rm UHE}\sim L_{\rm min}$,
which could be the case if the UHECR acceleration spectrum is hard,
$\alpha\lesssim2$. From Eqs.~(\ref{eq:flare}) and~(\ref{eq:delta_t}) we then have
\begin{equation}\label{eq:R_flare}
  R_{\rm f}\lesssim5\times10^{-3}\,\Gamma^4\,A^4\,Z^2\,
  \left(\frac{L_{\rm min}}{L_{\rm UHE}}\right)
  \left(\frac{E_{\rm max}}{10^{20}\ {\rm eV}}\right)^{-3}
  \,{\rm yr}^{-1}\,.
\end{equation}
In the limit $L_{\rm UHE}\to\overline{L_{\rm UHE}}$ we obviously recover the
limit of continuous sources, $R_{\rm f}\delta t\to1$.

During one flare the total non-thermal energy release would be
\begin{equation}\label{eq:R_flare}
  E_{\rm f}\gtrsim L_{\rm min}\delta t\gtrsim3\times10^{51}\,
  \Gamma^{-4}\,A^{-4}\,Z^{-2}\,
  \left(\frac{E_{\rm max}}{10^{20}\ {\rm eV}}\right)^5\,{\rm erg}\,,
\end{equation}
If this energy release is due to accretion onto a central black hole with
an energy extraction efficiency of $\sim10$\%~\cite{thorne74}, this corresponds to about
$0.02\,\Gamma^{-4}\,A^{-4}\,Z^{-2}\,M_\odot$. In AGN scenarios involving
accretion, this energy requirement is certainly modest.

Individual sources would be observed with the apparent UHECR luminosity
\begin{equation}\label{eq:R_flare}
  L_{\rm UHE,obs}\simeq L_{\rm UHE}\frac{\delta t}{t_{\rm delay}}\,,
\end{equation}
where $t_{\rm delay}\gtrsim Z^2\,10^5\,$yr is the time delay of charged cosmic
rays due to deflection in cosmic magnetic fields.

TeV $\gamma-$rays may or may not be observable from individual sources because
the duty cycle is small and most of the time the source would have luminosities
$\ll10^{45}\,Z^{-2}\,{\rm erg}\,{\rm s}^{-1}$, leading to fluxes
$\ll2\times10^{-8}\,Z^{-2}\,(d/20\,{\rm Mpc})^{-2}\,{\rm erg}\,{\rm cm}^{-2}\,{\rm s}^{-1}$
where $d$ is the distance to the source.
In the active phases, Eq.~(\ref{eq:tau_gg}) suggests that
TeV $\gamma-$ray may be absorbed by pair production within the sources. Note that
flares in the electromagnetic luminosity are not expected to correlate with
the UHECR luminosities due to the large UHECR time delays.

In the flaring limit we can have $L_{\rm UHE}\sim L_{\rm min}$ in which case
both $\gamma-$ray fluxes and secondary neutrino fluxes can not be much larger
than the UHECR flux which would then be a considerable fraction of the total energy budget.
Since the spectrum must be rather hard in this case, both the diffuse and discrete
neutrino fluxes are likely unobservably small.

\section{Centaurus A and other AGN Sources}
Centaurus A is the nearest AGN at a distance $d\simeq4\,$Mpc with a central
supermassive black hole of mass $M\sim10^8M_\odot$~\cite{CenA-review}.
It is not a blazar
as its jet has a large inclination angle to the line of sight.
The Pierre Auger Observatory measured two UHECR events from
the direction of Centaurus A~\cite{Cronin:2007zz}. This corresponds
to a flux~\cite{Cuoco:2007qd}
\begin{equation}\label{eq:j_CenA}
  \frac{dN_{\rm CR}}{dE}(E\simeq6\times10^{19}\,{\rm eV})\lesssim
  10^{-40}\,{\rm cm}^{-2}\,{\rm sr}^{-1}\,{\rm s}^{-1}\,{\rm eV}^{-1}\,,
\end{equation}
and to an apparent UHECR luminosity of Centaurus A of
$L_{\rm UHE,obs}\lesssim10^{39}\,{\rm erg}\,{\rm s}^{-1}$. If
Cen A is a continuous UHECR source, this is consistent with Eq.~(\ref{eq:Luhecr}).
The bolometric luminosity of Cen A is $L_{\rm tot}\simeq10^{44}\,{\rm erg}\,{\rm s}^{-1}$
which originates mostly within $\simeq500\,$pc from the center and is
mostly emitted around energies
$\varepsilon\sim1\,$eV~\cite{Quillen:2006qb,CenA-review}. This is
consistent with Eq.~(\ref{eq:Lmin}) if predominantly heavier nuclei are
accelerated, $Z\gtrsim4$. At MeV energies Cen A has a luminosity
$\simeq10^{42}\,{\rm erg}\,{\rm s}^{-1}$~\cite{CenA-review}.

If Cen A is emitting UHECR continuously, and assuming the UHECR injection spectrum
is $\propto E^{-\alpha}$, the expression analog to Eq.~(\ref{eq:j_inj_nu}) for
a discrete source gives for the secondary neutrino flux
\begin{equation}\label{eq:j_CenA_nu}
  j_\nu(E)\simeq\frac{2f}{3x_\nu}
  \frac{dN_{\rm CR}}{dE}(E/x_\nu)\,.
\end{equation}
Using Eq.~(\ref{eq:j_CenA}), one obtains numerically for the neutrino energy
flux
\begin{equation}\label{eq:j_CenA_nu2}
  E^2j_\nu(E)\simeq0.24\,x_\nu^{\alpha-1}f
  \left(\frac{E}{6\times10^{19}\,{\rm eV}}\right)^{2-\alpha}\,
  {\rm eV}\,{\rm cm}^{-2}\,{\rm s}^{-1}\,.
\end{equation}
For $\alpha\simeq2.2$ this gives
\begin{equation}\label{eq:j_CenA_nu3}
  E^2j_\nu(E)\simeq0.024\,f
  \left(\frac{E}{10^{17}\,{\rm eV}}\right)^{-0.2}\,
  {\rm eV}\,{\rm cm}^{-2}\,{\rm s}^{-1}\,,
\end{equation}
where from Eq.~(\ref{eq:f}), $f\lesssim100$. For $\alpha\simeq2.6$ it yields
\begin{equation}\label{eq:j_CenA_nu3}
  E^2j_\nu(E)\simeq0.37\,f
  \left(\frac{E}{10^{16}\,{\rm eV}}\right)^{-0.6}\,
  {\rm eV}\,{\rm cm}^{-2}\,{\rm s}^{-1}\,,
\end{equation}
where from Eq.~(\ref{eq:f}), $f\lesssim3.2$.
The latter, more optimistic case would give an event rate of
$\simeq1.3\times10^{-3}\,f\,{\rm yr}^{-1}\,{\rm km}^{-3}\lesssim
4.2\times10^{-3}\,{\rm yr}^{-1}\,{\rm km}^{-3}$. It is clear that
the rate due to the diffuse flux, estimated below Eq.~(\ref{eq:R_flare}),
is always much larger as long as CeN A is an "average" source. This
is consistent with the conclusion in Ref.~\cite{Koers:2008hv}.

If Cen A is an episodic UHECR source, as may be suggested by its
variability on time scales of days observed in X-rays and
$\gamma-$rays~\cite{CenA-review}, Eqs.~(\ref{eq:R_flare})
and~(\ref{eq:delta_t}) imply for the UHECR luminosity during a flare,
\begin{equation}\label{eq:CenA}
  L_{\rm UHE}\lesssim10^{45}\,\Gamma^4\,A^4\,Z^2\,
  \left(\frac{t_{\rm delay}}{Z^2\,10^5\,{\rm yr}}\right)
  \left(\frac{E_{\rm max}}{10^{20}\ {\rm eV}}\right)^{-3}
  \,{\rm erg}\,{\rm s}^{-1}\,.
\end{equation}
Comparing with Eq.~(\ref{eq:Lmin}) this suggests that the UHECR flare
luminosity is comparable to the total output, as long as the flare duration
is not much larger than the theoretical minimal variability time scale
Eq.~(\ref{eq:delta_t}).

The closest blazars whose jets are close to the line of sight and thus
may have considerably beamed emission are in general too far away to be
responsible for UHECR. As an example we briefly discuss  Markarian 501.
This blazar at a distance $d\simeq130\,$Mpc shows emission up to TeV
energies with variability on time scales
of days and peak luminosities of close to
$10^{46}\,{\rm erg}\,{\rm s}^{-1}$~\cite{Konopelko:2003zr,Kataoka}.
This is consistent with
Eq.~(\ref{eq:Lmin}) even for protons, $Z=1$. The power of such blazars
is, therefore, certainly sufficient to provide the UHECRs. Even if they
produce UHECR only in flares, the flare luminosity in UHECR, $L_{\rm UHE}$,
could be a small fraction of the total flare luminosity, and the necessary
flaring time scale Eq.~(\ref{eq:delta_t}) and rate Eq.~(\ref{eq:R_flare}) would
be consistent with observations, especially for significant beaming factors
$\Gamma$ typical for blazars.

\section{Conclusions}
We have discussed some consequences of latest results on
ultra-high energy cosmic rays for the nature and variability of the sources
as well as for the secondary
$\gamma-$ray and neutrino fluxes produced within the sources. To this end
we assumed predominant acceleration of nuclei of atomic mass $A$ and charge
$Z$. In the limit of continuously emitting sources,
their luminosity in cosmic rays above $\simeq6\times10^{19}\,$eV can be no
more than a fraction of $\simeq5\times10^{-4}\,Z^{-2}$ of the total source power.
If these cosmic rays are produced in the accretion disks in the centers
of AGNs, significant neutrino fluxes could be produced by hadronic interactions,
especially in scenarios in which extragalactic protons dominate down to
$\simeq10^{17}\,$eV such that the ankle is due to pair production of these protons.
The resulting cosmological diffuse neutrino flux can lead to detection
rates up to several events per year and km$^3$ of effective detection volume.
This also implies considerable photon fluxes at energies up to $\sim100\,$GeV,
the latter of which should be easily visible by GLAST. In contrast, TeV $\gamma-$rays are
likely absorbed within the source. For episodic sources that are beamed
by a Lorentz factor $\Gamma$, individual flares have to last at least
$\simeq0.1\,\Gamma^{-4}\,A^{-4}\,$yr. Such flares can also be visible
in photons up to the TeV energy range. A considerable fraction of the flare
luminosity could go into highest energy cosmic rays which suggests a hard
injection spectrum. In this case the rate of flares per source has to be
$\lesssim5\times10^{-3}\,\Gamma^4\,A^4\,Z^2\,{\rm yr}^{-1}$.
In contrast to continuously emitting sources, both neutrino fluxes from individual
sources and the resulting cosmological diffuse flux may be hard to detect in the
limit of flaring sources. Conversely, if high energy neutrinos are soon detected,
this may suggest sources that produce ultra-high energy cosmic rays continuously.

\section*{Acknowledgements}
I would like to thank Peter Biermann, Michael Kachelriess, Jim Matthews, Rachid Ouyed,
Georg Raffelt, Dmitry Semikoz, and Leo Stodolsky for useful discussions. I acknowledge
the Max Planck Institut f\"ur Physik in Munich for financial support of a visit
during which part of this paper was written.

\end{document}